\documentclass[cameraready]{Interspeech}

\title{Explainable and Trustworthy Speech Emotion Recognition Using Confidence Score and Reinforcement Learning Rectified Speech Emotion Descriptors}

\author[affiliation={1}]{Youjun}{Chen}
\author[affiliation={2*}]{Xurong}{Xie}
\author[affiliation={3}]{Mengzhe}{Geng}
\author[affiliation={4}]{Zengrui}{Jin}
\author[affiliation={1}]{Jiajun}{Deng}
\author[affiliation={1}]{Guinan}{Li}
\author[affiliation={1}]{Shujie}{Hu}
\author[affiliation={1}]{Huimeng}{Wang}
\author[affiliation={1}]{Haoning}{Xu}
\author[affiliation={1}]{Chengxi}{Deng}
\author[affiliation={1}]{Bowen}{Zhang}
\author[affiliation={1*}]{Xunying}{Liu}

\address{
    $^1$ The Chinese University of Hong Kong, Hong Kong SAR, China \\
    $^2$ Institute of Software, Chinese Academy of Sciences, China \\
    $^3$ National Research Council Canada, Canada \\
    $^4$ Tsinghua University, China
}

\email{\{yjchen, xyliu\}@se.cuhk.edu.hk, xurong@iscas.ac.cn}


\usepackage{comment}
\usepackage{tabularx}
\usepackage{multirow}
\usepackage{utfsym}
\usepackage{graphicx}
\usepackage{caption}
\usepackage{subcaption}

\usepackage{tabularx} 
\newcolumntype{Y}{>{\centering\arraybackslash}X} 

\begin{document}

\maketitle
\renewcommand{\thefootnote}{*}%
\footnotetext{Corresponding author.}%
\renewcommand{\thefootnote}{\arabic{footnote}}%
\begin{abstract}
Explainable and trustworthy speech emotion recognition (SER) remains a challenging task to date, largely due to the scarcity of SER data with reliable speech emotion descriptor (SED) labels, such as prosodic features and speaker traits. This paper presents a confidence score and reinforcement learning (RL) based on-the-fly SED rectification approach for post-training SER systems on automatically annotated SED labels. Experiments on IEMOCAP and MELD suggest that explainable SER systems incorporating the proposed confidence score and RL-based SED rectification approach consistently outperform baselines without data selection or SED rectification. The best performing system, which integrates both components, surpasses the baseline without data selection and SED rectification, achieving SER gains of 2.9\% and 3.3\% absolute (3.7\% and 5.4\% relative) on IEMOCAP and MELD benchmarks, respectively.
\end{abstract}

\section{Introduction}
Speech emotion recognition (SER) plays an important role in human-computer interaction. Despite decades of promising advancements, most traditional SER approaches \cite{tzirakis2018end,wang2020speech,shen2024emotion,zhang2024mersa,ma2024emotion2vec,tzeng2025noise} are typically limited to classifying speech into discrete emotion categories, while lacking the granularity and explainability necessary for a deeper and more human-like understanding of emotional expression \cite{xu2024secap}.

To this end, the development of large language models (LLMs) and speech-LLM models (SLMs) has facilitated the emergence of explainable SER tasks. Early explainable SER systems \cite{feng2024foundation,feng2024affect,li2025revise,yang2024large} first convert speech into transcribed text, and then apply LLMs to perform emotion recognition on the resulting transcription. Recent SLMs \cite{chen2025towards,santoso2024large,wang2025opens2s,mai2025aa,dutta2025leveraging} have notably improved both the accuracy and interpretability of SER by jointly leveraging speech-derived semantic information and speech emotion descriptor (SEDs) including prosodic features (e.g., pitch, volume) and speaker traits (e.g., gender, age). With advances in rule-based reinforcement learning, recent systems \cite{chen2025grpo,li2025emo,sun2025mecap,wang2026emotionthinker} have integrated novel group relative policy optimization (GRPO) method \cite{shao2024deepseekmath} into explainable SER, to further enhance SER performance using emotion recognition rewards.

However, these prior studies suffer from the following limitations: \textbf{a)} Lack of reliable SER data with ground-truth SED labels for further high quality supervised fine-tuning (SFT). Currently, due to the scarcity of SER data with reliable SED labels (e.g., expert annotations), previous methods \cite{chen2025towards,santoso2024large,wang2025opens2s,mai2025aa,dutta2025leveraging,chen2025grpo,li2025emo,sun2025mecap,wang2026emotionthinker} have largely relied on automated tools with a unified threshold for annotation. However, this approach yields relatively low reliability, as optimal threshold settings vary significantly across speakers of different ages, genders, and accents.
\textbf{b)} Lack of the ability to rectify unreliable SED labels during SER system training. Since SED labels are automatically generated offline with a unified threshold and not updated during training, these systems \cite{chen2025towards,santoso2024large,wang2025opens2s,mai2025aa,dutta2025leveraging,chen2025grpo,li2025emo,sun2025mecap,wang2026emotionthinker} are unable to rectify potentially incorrect labels based on the target emotional labels.
\textbf{c)} Lack of trustworthiness in the explainability of SER systems trained on automatically annotated SED labels without additional selection or rectification. Trained on unreliable SED labels, prior SER systems \cite{chen2025towards,santoso2024large,wang2025opens2s,mai2025aa,dutta2025leveraging,chen2025grpo,li2025emo,sun2025mecap,wang2026emotionthinker} may produce inaccurate SED predictions, even if the ultimate SER result is correct. Therefore, explainable and trustworthy SER remains a challenging task to date.

To this end, this paper proposes a confidence score and RL-based on-the-fly SED rectification approach for post-training SER systems on automatically annotated SED labels. First, we train an MLP-based confidence estimation model (CEM) to evaluate the confidence scores of automatically annotated SED labels, which is then leveraged to select a more reliable subset for further SFT. Second, during SER system training, an RNN-based SED Controller is alternatively updated to generate on-the-fly SED rectification policies. With an appropriate data selection ratio, the combination of data selection and on-the-fly SED rectification enables explainable SER systems to achieve both enhanced trustworthiness and improved SER performance.

Experiments on the IEMOCAP and MELD corpus suggest that explainable SER systems incorporating the proposed confidence score and RL-based SED rectification approach consistently outperform baselines without data selection or SED rectification. The best performing system, which integrates both components, surpasses the baseline trained on the full data without data selection and SED rectification, achieving SER gains of 2.9\% and 3.3\% absolute (3.7\% and 5.4\% relative) on IEMOCAP and MELD benchmarks, respectively. Extensive experiments on different model configurations and parameter settings further demonstrate that these two components jointly and consistently enhance SER performance and trustworthiness.

The main contributions of our work are summarized below:

\textbf{1)} This paper pioneers the first work to leverage the confidence score-based data selection method for SER data with automatically annotated SED labels, with the goal of selecting a reliable subset for further high quality SFT.

\textbf{2)} This paper presents the first use of a RL-based SED Controller to generate on-the-fly SED rectification policies targeting explainable and trustworthy SER.

\textbf{3)} To the best of our knowledge, this is the first work to systematically explore the impact of SED label quality on the overall performance of explainable SER systems. Extensive comparative experiments and t-SNE visualization demonstrate that reliable SED labels significantly enhance both the SER performance and the trustworthiness of the explainable SER systems.

\begin{figure*}[htbp]
\centering
\setlength{\abovecaptionskip}{0pt plus 1pt minus 3pt}
\includegraphics[scale=0.16]{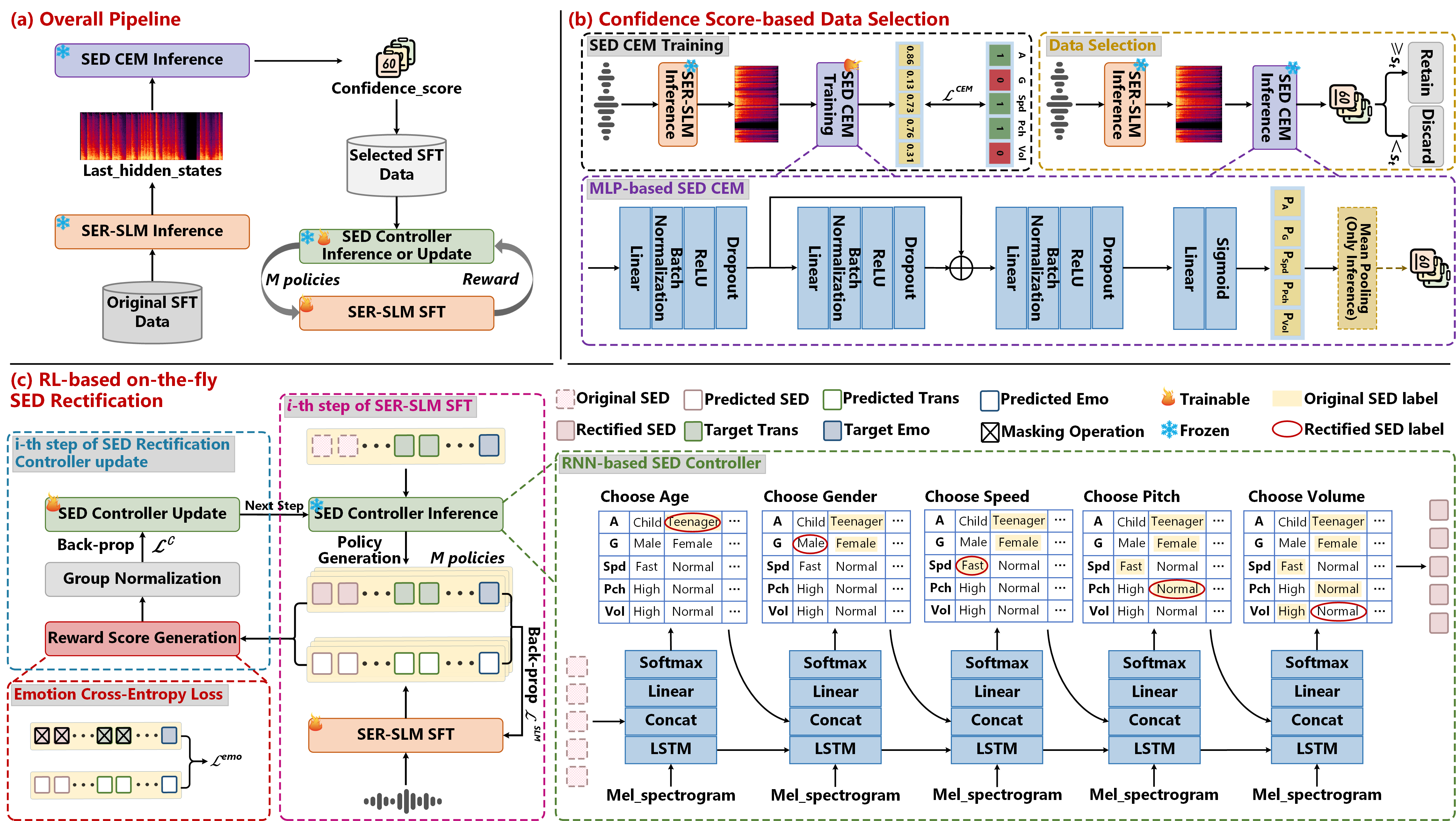}
\caption{Illustration of (a) an overall pipeline of the proposed confidence score and RL based on-the-fly SED rectification post-training approach for explainable and trustworthy SER systems, including (b) confidence score-based data selection; and (c) RL-based on-the-fly SED rectification. Furthermore, confidence-based data selection and RL-based SED rectification are optional components of the overall pipeline. The abbreviations ``A'', ``G'', ``Spd'', ``Pch'' and ``Vol'' denote age, gender, speed, pitch, and volume, respectively.}
\label{fig1}
\vspace{-0.6cm}
\end{figure*}

\section{Confidence Score-based Data Selection}
\vspace{-0.1cm}
Confidence scores quantify prediction reliability by expressing outputs as continuous probabilities, mitigating the overconfidence inherent in discrete classifications \cite{deng2023confidence}.
They have been widely adopted in speech recognition \cite{ragni2022increasing,li2021confidence,kahn2020self,pan2021using}, speaker adaptation \cite{deng2023confidence,gollan2008confidence,deng2022confidence} and audio generation \cite{zhu2025cosyaudio,borsos2023soundstorm}. To mitigate the sensitivity of SER system performance to supervision error rates during post-training, this research explores a confidence score-based data selection approach to select a more reliable subset based on SED confidence scores for further SFT. 

\vspace{-0.2cm}
\subsection{Confidence estimation module}
\vspace{-0.1cm}
We adopt the MLP-based confidence estimation model (CEM) from \cite{deng2023confidence} as the backbone of our CEM, and extend it by adding a Mean Pooling layer to aggregate confidence scores of individual SED labels into a single confidence score. As shown in Figure \ref{fig1}(b), our CEM is a lightweight binary classification
model, which contains a simple yet effective 3-layer residual
feed-forward network and a
Sigmoid and Mean Pooling layer. Batch normalization, ReLU activation, and dropout neural operations are applied to each hidden layer in sequence. Skip connection is also utilized between
the outputs of the first and the second hidden layer. This CEM is then connected with the pre-trained SER-SLM system. For each utterance, the hidden states from the last LLM decoder layer of the SLM, namely last hidden states, before the final LM head are fed into the CEM as input features, which then outputs a smoothed SED confidence score by averaging the confidence scores for the individual SED predictions, namely pitch, volume, speed, gender, and age.

\vspace{-0.2cm}
\subsection{CEM training and data selection}
\vspace{-0.1cm}
To produce utterance-level confidence score labels for CEM training, we perform inference using a pre-trained SER-SLM system on SER data with reliable SED labels. Specifically, for each utterance $d$ in a given dataset $\mathcal{D}$ with ground-truth SED labels $[a_{gt}, g_{gt}, s_{gt}, p_{gt}, v_{gt}]$, we obtain both the predicted SED array $[a_{pred}, g_{pred}, s_{pred}, p_{pred}, v_{pred}]$ and the corresponding last hidden states $\textbf{\textit{h}}_d$. For each SED prediction, a binary label is assigned: one denotes a correct prediction and zero an incorrect one, yielding a five-dimensional binary array $\textbf{\textit{l}}_d$. During CEM training, we first set aside the Mean Pooling layer, then take $\textbf{\textit{h}}_d$ as input and $\textbf{\textit{l}}_d$ as supervision label. The following cross-entropy loss function computed between the five-dimensional confidence score array $\textbf{\textit{c}}_d$ and the target label $\textbf{\textit{l}}_d$ is used to optimize the CEM.

\vspace{-0.4cm}
\begin{align}
\mathcal{L}^{CEM} = \sum_{d\in \mathcal{D}}(\textbf{\textit{l}}_d\log(\textbf{\textit{c}}_d)+(\textbf{1}-\textbf{\textit{l}}_d)\log(\textbf{1}-\textbf{\textit{c}}_d))
\label{eq_cemloss}
\vspace{-0.1cm}
\end{align}

For data selection, we use the same SER-SLM to extract $\textbf{\textit{h}}_g$ from each sample $g$ in the target dataset $G$, yielding a corresponding confidence score array $\textbf{\textit{c}}_g$. Five scores in $\textbf{\textit{c}}_g$ are then aggregated via the Mean Pooling output layer to produce the utterance-level comprehensive confidence score $s_g$, which is used to determine whether to retain or discard this sample. By presetting a threshold $s_t$, we can flexibly control the proportion of selected data, forming the selected SFT data $G^\prime$.

\vspace{-0.2cm}
\section{RL-based SED Controller Rectification}
\vspace{-0.1cm}
Although confidence score-based data selection is an effective method, a low data retention rate introduces two extra challenges: the filtering of difficult samples and the reduction of training data. To address this, we propose a RL-based approach for the automatic, on-the-fly SED rectification in SER-SLM training, where an alternating update procedure is applied, drawing on the policy optimization method used in \cite{jin2024towards}.

\vspace{-0.2cm}
\subsection{RNN-based SED Controller}
\vspace{-0.1cm}
As illustrated in the green box of Figure \ref{fig1}(c), the RNN-based SED Controller, parameterized by $\bm{\theta}^C$, consists of a one-layer LSTM module, followed by a Concat layer, a Linear layer and a Softmax decoder layer. The mel-spectrogram of the input speech is first passed through an LSTM layer to obtain compressed representations, which are then concatenated with the embeddings of the original SED tokens and fed into the final Linear and Softmax layers. For each SED label, the SED rectification process chooses between two operations: retaining or modifying, which is decided based on the corresponding probability predicted by the SED Controller. Finally, we obtain the rectified SED tokens $\textbf{\textit{Y}}^{sed}$ as the supervision labels.

\vspace{-0.2cm}
\subsection{SED rectification and model update}
\vspace{-0.1cm}
\noindent \textbf{SER-SLM SFT:} As illustrated in the pink box of Figure \ref{fig1}(c), at the $i$-th training step, the SED Controller, with its parameters frozen, first generates $M$ SED rectification policies $\prod_i=[\pi_{i,1},\pi_{i,2},\cdots,\pi_{i,M}]$. Next, a structured supervision label $\textbf{\textit{Y}}_i$ is constructed for training SER-SLM $\bm{\theta}^{SLM}$, comprising the ground-truth transcription $\textbf{\textit{Y}}_{i}^{trans}$, emotional labels $\textbf{\textit{Y}}_{i}^{emo}$, and rectified SED labels $\textbf{\textit{Y}}_{i}^{sed}$. For each speech utterance with speech and text prompt $\textbf{\textit{X}}_i$, there are $M$ such possible supervision labels. The overall loss function $\mathcal{L}_i^{SLM}$ for the SER-SLM update is computed by aggregating and averaging the LLM loss across all $M$ sampled SED rectification policies as follows:

\vspace{-0.5cm}
\begin{align}
\mathcal{L}_i^{SLM} = \frac{1}{M}\sum_{m=1}^M\mathcal{L}^{LLM}_{i,m}(\textbf{\textit{X}}_i,\pi_{i,m},\bm{\theta}_{i-1}^{SLM},\textbf{\textit{Y}}_i)
\label{eq_SLMloss}
\end{align}

\vspace{-0.1cm}
\noindent \textbf{Reward function:} Following each SER-SLM training step, the corresponding policy reward $R$ is calculated based on the performance changes brought about by different SED rectification policies, which in turn serves as a guidance signal for updating the SED Controller. As shown in the red box of Figure \ref{fig1}(c), we mask all tokens except ground-truth emotion token in labels, and compute the emotion corresponding cross-entropy loss $\mathcal{L}^{emo}$, following Equation \ref{eq_emoloss}. A total of $M$ losses are collected and then normalized to compute the policy reward $R$.

\vspace{-0.3cm}
\begin{align}
R(\bm{\theta}_{i-1}^{C},\bm{\theta}_{i}^{SLM})&=\mathbb{E}_{p(\prod_i|\prod_{i-1:1};\bm{\theta}_{i-1}^{C})}[-\mathrm{norm}(\mathcal{L}_{i}^{emo})]
\label{eq_reward}
\end{align}


\vspace{-0.5cm}
\begin{align}
\mathcal{L}_{i}^{emo}=-\log p\big(\textbf{\textit{Y}}_{i}^{emo}|(\textbf{\textit{Z}}_{i}^{sed},\textbf{\textit{Z}}_{i}^{trans});\bm{\theta}_{i}^{SLM}\big)
\label{eq_emoloss}
\end{align}
where, $\textbf{\textit{Z}}_{i}^{sed}$ and $\textbf{\textit{Z}}_{i}^{trans}$ denote the predicted SED and Transcription tokens, respectively. $\mathrm{norm}(\cdot)$ is the group normalization operation here.

\noindent \textbf{SED Controller update:} In the $i$-th SED Controller update phase, its parameters become trainable. The training objective of SED Controller is computed by averaging over the sampled $M$ rectification policies, following Equation \ref{eq_controllerloss}.

\vspace{-0.3cm}
\begin{align}
&\nabla_{\bm{\theta}^C}\mathcal{L}_i^C = \nabla_{\bm{\theta}^C}[- R(\bm{\theta}_{i-1}^{C},\bm{\theta}_{i}^{SLM})] \nonumber \\
& =\nabla_{\bm{\theta}^C}\sum_{\pi_{i,m}}p(\pi_{i,m}|\pi_{i-1:1,m};\bm{\theta}_{i-1}^{C})\cdot\mathrm{norm}(\mathcal{L}_{i}^{emo})] \nonumber \\
&\approx \frac{1}{M}\sum_{m=1}^M\mathrm{norm}(\mathcal{L}_{i}^{emo}) \cdot \nabla_{\bm{\theta}^C}p(\pi_{i,m}|\pi_{i-1:1,m};\bm{\theta}_{i-1}^{C})
\label{eq_controllerloss}
\end{align}

\vspace{-0.2cm}
\noindent \textbf{Implementation issues: 1) Reward normalization:} Similar to GRPO, the $M$ emotion cross-entropy rewards at the i-th epoch are mean and variance normalized across $M$ policies to ensure the stable convergence of the SED Controller update. The gradient of SED Controller is propagated via the probability of generating a specified policy under the new policy model conditioned on the old one, scaled by the group normalized policy reward.

\noindent \textbf{2) Number of policy samples:} An ablation study on the performance comparison using varying numbers of policy samples during RL-based on-the-fly SED rectification is presented later in Table \ref{tab_ablation}. Empirically the number of policy samples is set as $M=6$ unless otherwise stated.

\noindent \textbf{3) Policy space definition:} For each automatically annotated SED label, the SED Controller decides whether to retain or modify it based on the predicted probability.

\vspace{-0.3cm}
\section{Overall Pipeline}
\vspace{-0.1cm}
As illustrated in Figure \ref{fig1}(a), the overall pipeline of the proposed confidence score and RL based on-the-fly SED rectification approach first implement confidence score based data selection, followed by on-policy SED rectification and SER-SLM post-training on the selected data. 

\begin{table*}[ht]
\centering
\caption{SER performance comparison with baselines on IEMOCAP and MELD test data. The column ``SED" indicates whether the SER system incorporates SED information. “Domain SFT” denotes whether the pre-trained SER-SLM is further fine-tuned on the IEMOCAP and MELD training sets. ``$\ast$'' and ``$\ddagger$'' represent statistically significant (Paired Single-tailed t-TEST \cite{champoux2022first}, $p$=0.05) accuracy improvement over Sys. 4 and Sys. 5, respectively.}
\vspace{-0.1cm}
\label{tab_main}
\setlength{\tabcolsep}{3.2pt} 
\fontsize{9}{10}\selectfont 
\begin{tabularx}{\textwidth}{c|c|c|c|c|c|c|c}
\toprule
\textbf{Sys} & \textbf{SED} & \textbf{Domain SFT} & \textbf{Confidence Score Data Selection} & \textbf{RL-based SED Rectification} & \textbf{IEMOCAP\%} & \textbf{MELD\%} & \textbf{Avg\%} \\
\midrule
\hline
1 & \multirow{2}{*}{$\usym{2717}$} & $\usym{2717}$ & \multirow{4}{*}{$\usym{2717}$} & \multirow{8}{*}{$\usym{2717}$} & 63.01 & 48.73 & 53.33 \\
\cline{1-1} \cline{3-3} \cline{6-8}
2 & & $\usym{2713}$ & & & 73.65 & 56.29 & 61.89 \\
\cline{1-3} \cline{6-8}
3 & \multirow{11}{*}{$\usym{2713}$} & $\usym{2717}$ & & & 65.19 & 51.53 & 55.93 \\
\cline{1-1} \cline{3-3} \cline{6-8}
4 & & \multirow{10}{*}{$\usym{2713}$ }& & & 78.08 & 60.81 & 66.38 \\
\cline{1-1} \cline{4-4} \cline{6-8}
5 &  & & 90\% & & 78.40 & 62.04 & 67.31 \\
\cline{1-1} \cline{4-4} \cline{6-8}
6 &  &  & 80\% & & 78.89 & 61.92 & 67.39 \\
\cline{1-1} \cline{4-4} \cline{6-8}
7 &  & & 70\% & & 77.68 & 60.81 & 66.25 \\
\cline{1-1} \cline{4-4} \cline{6-8}
8 &  & & 60\% & & 75.02 & 58.13 & 63.58 \\
\cline{1-1} \cline{4-6}\cline{6-8}
9 & & & $\usym{2717}$ & \multirow{6}{*}{$\usym{2713}$} & 79.85 & 62.96 & 68.41 \\
\cline{1-1} \cline{4-4} \cline{6-8}
10 & & & 90\% &  & \textbf{80.98}$\ast$$\ddagger$ & \textbf{64.11}$\ast$$\ddagger$ & \textbf{69.55}$\ast$$\ddagger$ \\
\cline{1-1} \cline{4-4} \cline{6-8}
11 & & & 80\% & & 80.66 & 63.77 & 69.45 \\
\cline{1-1} \cline{4-4} \cline{6-8}
12 & & & 70\% & & 78.24 & 61.46 & 66.87 \\
\cline{1-1} \cline{4-4} \cline{6-8}
13 & & & 60\% & & 75.66 & 58.51 & 64.04 \\
\hline
\bottomrule
\end{tabularx}
\vspace{-0.2cm}
\end{table*}

\begin{table}[ht]
\centering
\caption{SER performance comparison with SOTA open-source SER-SLM systems on IEMOCAP and MELD test data.}
\vspace{-0.1cm}
\label{tab_sota}
\setlength{\tabcolsep}{3.5pt} 
\fontsize{9}{10}\selectfont 
\begin{tabularx}{\columnwidth}{ccccc}
\toprule
\textbf{Sys} & \textbf{IEMOCAP\%} & \textbf{MELD\%} & \textbf{Avg\%} \\
\midrule
Kimi-Audio \cite{ding2025kimi} & 57.72 & 59.13 & 58.68 \\
Qwen2-Audio \cite{chu2024qwen2} & 37.71 & 51.23 & 46.87 \\
Audio-Flamingo-3 \cite{goel2025audio} & 69.06 & 56.71 & 60.69 \\
Step-Audio-R1 \cite{tian2025step} & 53.99 & 46.43 & 48.87 \\
OSUM-EChat \cite{geng2025osum} & 41.49 & 53.38 & 49.55 \\
BLSP-Emo \cite{wang2025opens2s} & 75.99 & 57.29 & 63.32 \\
VIB-Emo \cite{chen2025towards} & 77.60 & 60.12 & 65.76 \\
Ours (Sys. 10, Tab. \ref{tab_main}) & \textbf{80.98} & \textbf{64.11} & \textbf{69.55} \\
\bottomrule
\end{tabularx}
\end{table}

\begin{table}[ht]
\centering
\caption{Ablation study on the number $M$ of SED rectification policy samples generated by SED Controller, where $M \in \{2, 4, 6, 8, 10\}$, and Sys. V3 corresponds to Sys. 10 in Table \ref{tab_main}. }
\vspace{-0.1cm}
\label{tab_ablation}
\setlength{\tabcolsep}{12.3pt} 
\fontsize{9}{10}\selectfont 
\begin{tabularx}{\columnwidth}{ccccc}
\toprule
\textbf{Sys} & $\mathbf{M}$ & \textbf{IEMOCAP} & \textbf{MELD} & \textbf{Avg} \\
\midrule
V1 & 2 & 79.37 & 62.69 & 68.07 \\
V2 & 4 & 79.94 & 63.04 & 68.49 \\
V3 & 6 & 80.98 & \textbf{64.11} & \textbf{69.55} \\
V4 & 8 & \textbf{81.06} & 63.69 & 69.29 \\
V5 & 10 & 80.26 & 63.57 & 68.95 \\
\bottomrule
\end{tabularx}
\vspace{-0.5cm}
\end{table}

\vspace{-0.3cm}
\section{Experiments}
\vspace{-0.1cm}
\subsection{Experimental setup}
\vspace{-0.1cm}
\noindent \textbf{Baselines:} Our work builds upon the SER-SLM introduced in \cite{chen2025towards}, where variational information bottleneck (VIB) \cite{alemi2016deep} are applied to disentangle SED and semantic features from HuBERT embeddings. Moreover, we compare our system with four open-source general-purpose SLMs: Kimi-Audio \cite{ding2025kimi}, Qwen2-Audio \cite{chu2024qwen2}, Audio-Flamingo-3 \cite{goel2025audio}, and Step-Audio-R1 \cite{tian2025step}; as well as three explainable SER-focused SLMs: OSUM-EChat \cite{geng2025osum}, BLSP-Emo \cite{wang2025opens2s}, and VIB-Emo \cite{chen2025towards}.

\noindent \textbf{Datasets:} We use the same publicly available datasets as \cite{chen2025towards}, including large scale subset GigaSpeech-m in SpeechCraft with about 670k utterances \cite{jin2024speechcraft}, IEMOCAP \cite{busso2008iemocap} and MELD \cite{poria2019meld}. \textbf{1) Pre-training:} Given that SpeechCraft provides SED labels, transcriptions, and emotion labels, we use the following template to restructure the structured audio captions: \textit{``The \textless Age\textgreater\ \textless Gender\textgreater\ speaks with a \textless speed\textgreater, a \textless pitch\textgreater\ and a \textless volume\textgreater, saying `\textless transcription\textgreater'. Based on the audio caption, the emotional state of this speaker is \textless emotion\textgreater"}. Although SED labels are also generated through automated annotation, initializing the SER-SLM on a large amount of relatively reliable data accelerates convergence compared to pre-training from scratch. \textbf{2) CEM training:} Our CEM is trained on the same pre-training data used for the SER-SLM. For data balance, we first performed inference using the pre-trained SER-SLM on this data, obtaining 261.3k incorrect and 408.7k correct SED predictions, from which we randomly selected 100k positive-negative sample pairs for CEM training. \textbf{3) Post-training:} First, the tools and thresholds provided by SpeechCraft are used to automatically annotate SED labels on the train and development subsets in IEMOCAP and MELD, which serve as the original SED labels. We then perform post-training of the pre-trained SER-SLM system, leveraging the automatically annotated data. \textbf{4) Evaluation:} We evaluate different SER models on the same IEMOCAP and MELD test data as \cite{chen2025towards}.

\noindent \textbf{Training details: 1) SER-SLM training:} During SER-SLM pre-training and post-training, we use the AdamW optimizer with a learning rate of $2\times10^{-4}$ and a warmup ratio of $10\%$. We pre-train the system for two epochs and conduct post-training for 20k steps. \textbf{2) CEM training:} For CEM training, the Adam optimizer is applied with a learning rate of $1\times10^{-3}$, and the dropout ratio is set as 0.1. \textbf{3) SED Controller training:} The SED Controller has a hidden size of 128 and an embedding size of 32 using the Adam optimizer with a learning rate of $3\times10^{-4}$.

\begin{figure}[htbp]
    \centering
    \begin{subfigure}[b]{0.23\textwidth}
        \centering
        \includegraphics[width=\textwidth]{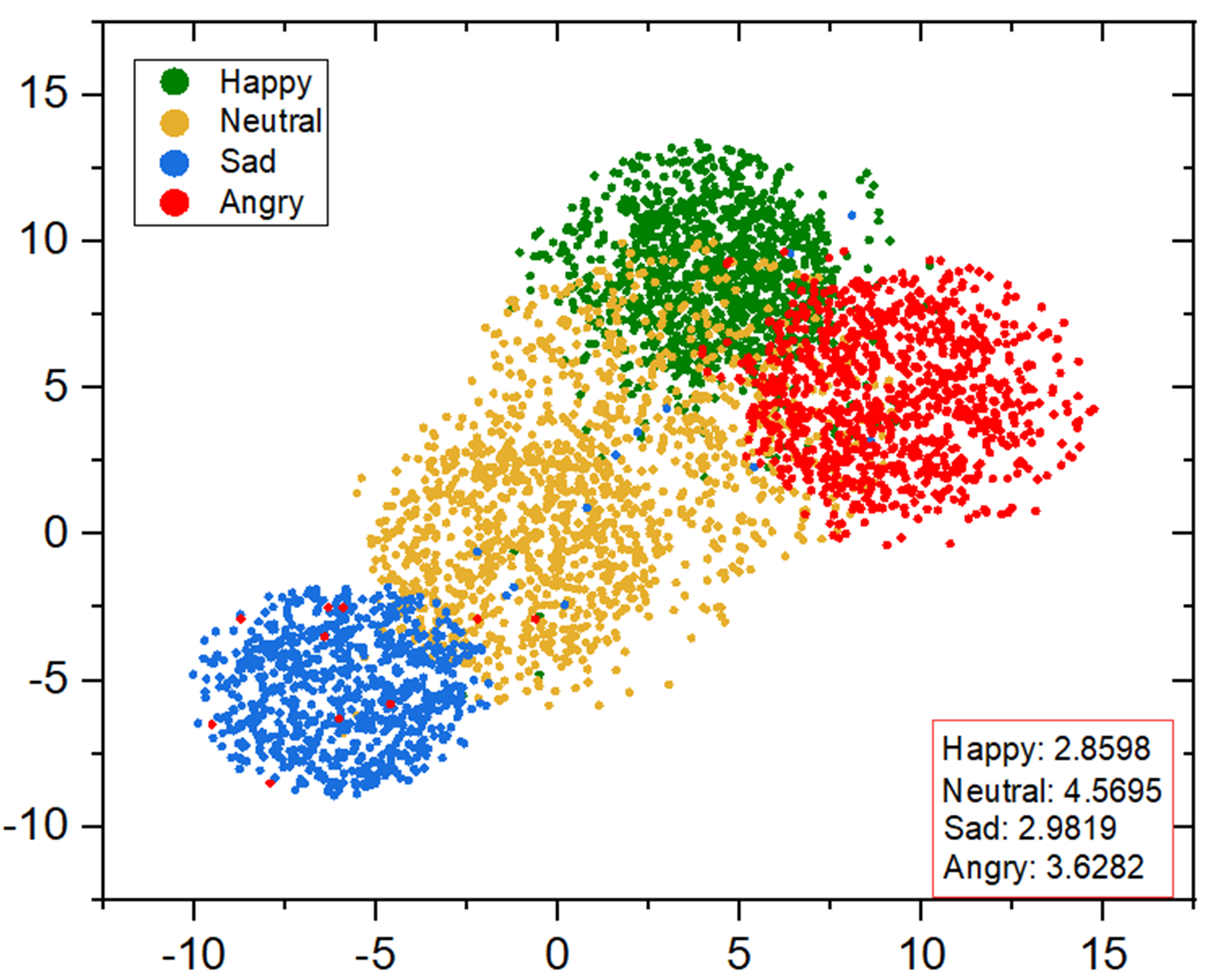}
        \caption{Baseline}
        \label{fig:subfig1}
    \end{subfigure}
    \hfill 
    \begin{subfigure}[b]{0.23\textwidth}
        \centering
        \includegraphics[width=\textwidth]{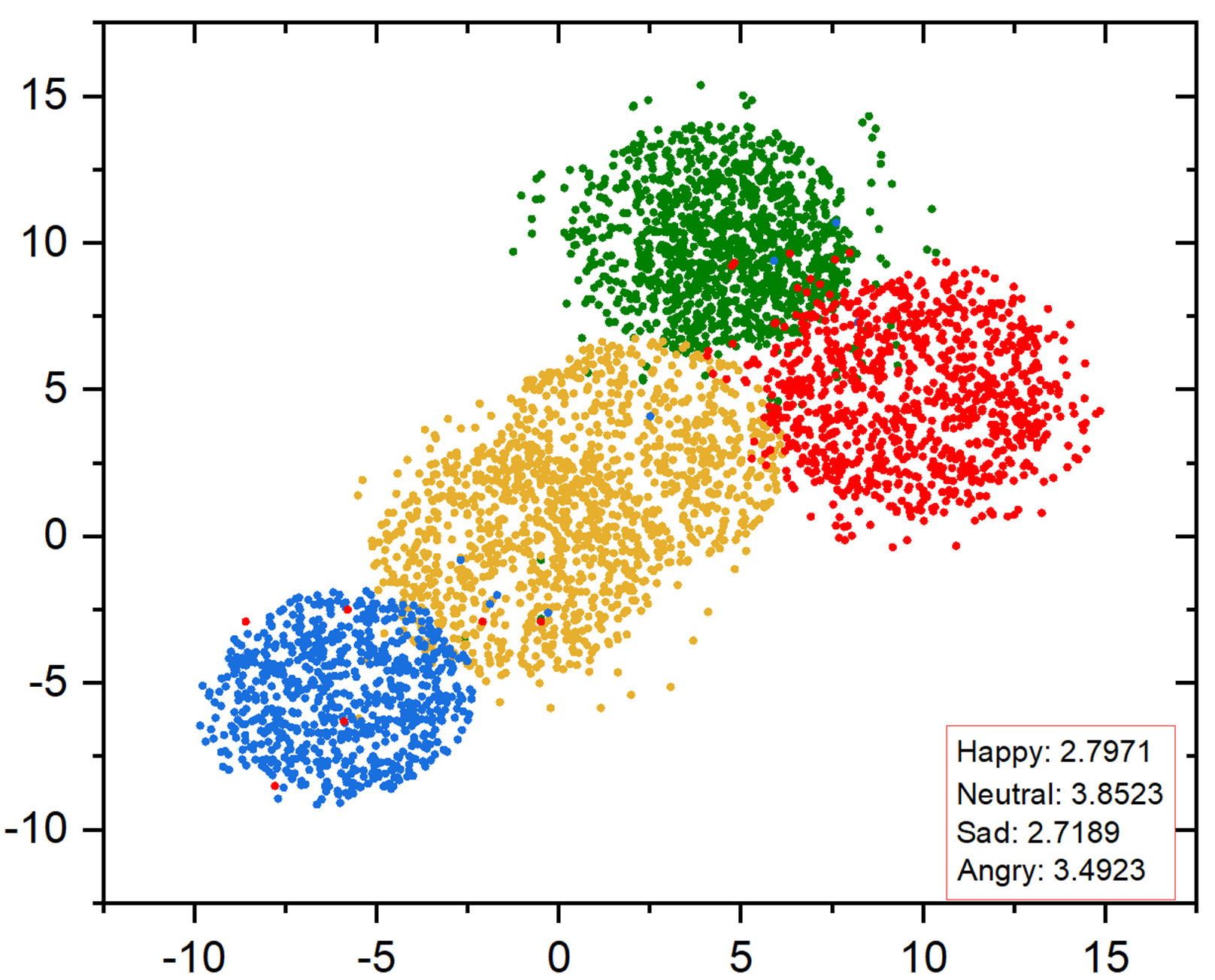}
        \caption{Best performing system}
        \label{fig:subfig2}
    \end{subfigure}
    \vspace{-0.2cm}
    \caption{t-SNE visualization of the last hidden states extracted from the baseline (Sys. 4, Table \ref{tab_main}) and the best performing system (Sys. 10, Table \ref{tab_main}) for IEMOCAP training set. In the red box (bottom right) of each graph, the covariance log determinants of four emotion clusters are listed.}
    \label{fig_tsne}
\vspace{-0.5cm}
\end{figure}

\vspace{-0.2cm}
\subsection{Main results}
\vspace{-0.1cm}
\noindent \textbf{Analysis on SER performance comparison with baselines and SOTA SER systems:} As shown in Table \ref{tab_main}, integrating SED information plays a crucial role in enhancing both the performance and explainability of SER systems (Sys. 3 vs. Sys. 1; Sys. 4 vs. Sys. 2). Moreover, explainable SER systems incorporating the proposed confidence score and RL-based SED rectification approach consistently outperform baselines that lack data selection or SED rectification (Sys. 10–13 vs. Sys. 3–9). The best performing system (Sys. 10), integrating both components, surpasses the baseline trained on the full dataset without data selection and SED rectification (Sys. 4), achieving SER gains of 2.9\% and 3.3\% absolute (3.7\% and 5.4\% relative) on IEMOCAP and MELD, respectively. In addition, while the optimal data selection ratio is 80\% when used alone (Sys. 6), it shifts to 90\% when combined with SED rectification (Sys. 10), indicating that a higher threshold is not necessarily optimal when SED labels can be on-policy rectified and tightly integrated with SER system training. Finally, the performance of the best performing SER system (Sys. 10, Table \ref{tab_main}) is further contrasted with the results using open-source SLMs in Table \ref{tab_sota}.

\noindent \textbf{Ablation study on the number of SED rectification policy samples:} As shown in Table \ref{tab_ablation}, the on-the-fly SED rectification approach with $M=6$ (Sys. V3) delivers the best overall SER performance, although it trails the optimal setting on the IEMOCAP test set by a narrow margin of 0.08\%.

\noindent \textbf{Visualization:} In Figure \ref{fig_tsne}, we present a t-SNE visualization of the last hidden states extracted from the baseline (Sys. 4, Table \ref{tab_main}) and the best performing system (Sys. 10, Table \ref{tab_main}) for IEMOCAP training set. The sharper cluster boundaries observed in subgraph (b) indicate that post-training on higher quality SED labels improves SER performance, especially in better distinguishing between emotions such as ``Neutral'' and ``Happy''.

\vspace{-0.3cm}
\section{Conclusion}
\vspace{-0.1cm}
This paper proposes a confidence score and RL-based on-the-fly SED rectification approach for post-training SER systems on automatically annotated SED labels. Experiments on IEMOCAP and MELD suggest that explainable SER systems incorporating the proposed confidence score and RL-based SED rectification approach consistently outperform baselines without data selection or SED rectification and other open-source SLMs.

\section{Acknowledgements}
This research is supported by Hong Kong RGC GRF grant No. 
14200021, 14200324, 
National Natural Science Foundation of China No. 62302494, Beijing Natural Science Foundation-Xiaomi Innovation Joint Fund No. L243035, and Youth Innovation Promotion Association CAS Grant No. 2023119.

\section{Generative AI Use Disclosure}
Generative AI tools were used only for language editing and improving readability during the preparation of this manuscript. These tools were not used to generate core scientific ideas, experimental data, or technical contributions. All authors have thoroughly reviewed and approved the final manuscript and assume full responsibility for the integrity of its entire content.

\bibliographystyle{IEEEtran}
\bibliography{mybib}

\end{document}